\documentclass[a4paper]{article}

\usepackage{INTERSPEECH2021}
\usepackage{IEEEtrantools}
\usepackage{url}
\usepackage{xcolor}
\usepackage{multirow}

\urldef{\voxsrc20}\url{http://www.robots.ox.ac.uk/~vgg/data/voxceleb/competition2020.html}
\urldef{\dscore}\url{https://github.com/nryant/dscore}
\urldef{\dihard3baseline}\url{https://github.com/dihardchallenge/dihard3_baseline/}

\title{The Third DIHARD Diarization Challenge}
\name{Neville Ryant$^{1}$,
      Prachi Singh$^{2}$,
      Venkat Krishnamohan$^{2}$,
      Rajat Varma$^{2}$,
      Kenneth Church$^{3}$,
      Christopher Cieri$^{1}$,
      Jun Du$^{4}$, 
      Sriram Ganapathy$^{2}$,
      Mark Liberman$^{1}$
    }

\address{
  $^1$Linguistic Data Consortium, University of Pennsylvania, Philadelphia, PA, USA \\
  $^2$LEAP Lab, Electrical Engineering, Indian Institute of Science, Bangalore, India, \\
  $^3$Baidu Research, Sunnyvale, CA, USA \\
  $^4$University of Science and Technology of China, Hefei, China. } 

\email{nryant@ldc.upenn.edu}

\begin{document}
\bstctlcite{IEEEexample:BSTcontrol}
\ninept
\maketitle


%
\begin{abstract}
    DIHARD III was the third in a series of speaker diarization challenges intended to improve the robustness of diarization systems to variability in recording equipment, noise conditions, and conversational domain. Speaker diarization was evaluated under two speech activity conditions (diarization from a reference speech activity vs. diarization from scratch) and $11$ diverse domains. The domains span a range of recording conditions and interaction types, including read audio-books, meeting speech, clinical interviews, web videos, and, for the first time, conversational telephone speech. A total of 30 organizations (forming 21 teams) from industry and academia submitted 499 valid system outputs. The evaluation results indicate that speaker diarization has improved markedly since DIHARD I, particularly for two-party interactions, but that for many domains (e.g., web video) the problem remains far from solved.
\end{abstract}

\section{Introduction}
Speaker diarization, often referred to as ``who spoke when'', is the task of determining the number of speakers  present in a conversation and their respective regions of activity in the given recording. In addition to being an interesting technical challenge, it forms an important part of the pre-processing pipeline for speech-to-text \cite{watanabe2020chime} and is essential for making objective measurements of turn-taking behavior. Early work in this area was driven by the NIST Rich Transcription (RT) evaluations \cite{fiscus2006rich}, which ran from 2002 to 2009. In addition to driving substantial performance improvements, especially for meeting speech, the RT evaluations introduced  diarization error rate (DER), which remains the principal evaluation metric in this area. 

After the RT evaluation series ended in $2009$, diarization systems continued to improve (e.g., i-vectors, x-vectors, PLDA scoring), though until quite recently there was no common benchmark for diarization, resulting in a fragmented research landscape where individual groups focused on different datasets or domains (e.g., conversational telephone speech \cite{sell2014speaker,zhu2016online,garcia2017speaker,wang2018speaker,zhang2019}, broadcast news \cite{rouvier2013open,vinals2017domain}, or meeting \cite{yella2013improved,yella2014artificial}), often with slightly differing evaluation methodologies. At best, this has made performance comparisons difficult,  while at worst it may have engendered over-fitting to individual domains/datasets, resulting in systems that do not generalize.

Recently, there has been renewed interest in a common diarization task to facilitate systematic bench marking. Whereas from 2009-2017 there were no major evaluations with a diarization component, there now is an annual diarization specific evaluation  --  DIHARD  --  as well as numerous other challenges that include a diarization component; among others, the Fearless Steps series \cite{hansen2019,joglekar2020fearless}, the Iberspeech-RTVE challenge \cite{lleida2019albayzin}, CHiME-6 \cite{watanabe2020chime}, and VoxSRC-20 \cite{nagrani2020voxsrc}..

The first DIHARD challenge (DIHARD I) \cite{dihard1_eval_plan} ran in the spring of 2018 and evaluated diarization of single channel wideband recordings drawn from a diverse range of domains. As expected, state-of-the-art systems performed poorly, with error rates of the best systems \cite{sell2018diarization,diez2018but} more than double the state-of-the-art for CALLHOME \cite{cieri2003switchboard} at the time \cite{garcia2017speaker,wang2018speaker}.  This was followed by DIHARD II \cite{dihard2_eval_plan,ryant2019second} in 2019, which was even more successful, attracting 50 teams from 17 countries and 4 continents. While DIHARD II continued the single channel diarization tracks from DIHARD I, it also collaborated with the CHiME challenge series with the addition of two new tracks focusing on conversational speech from multiple far-field microphone arrays during a dinner party scenario. All tracks continued to be challenging for participants, with the most challenge tracks being the ones without reference speech segmentation and dinner party data. In the case of the latter, the CHiME-6 data, DER of the best performing system was over 45\% when provided with an oracle speech segmentation and over 58\% when required to produce its own segmentation.

The current challenge (DIHARD III)\footnote{\url{https://dihardchallenge.github.io/dihard3/}} \cite{dihard3_eval_plan}, which builds upon DIHARD I and II, addresses the problem of robust diarization which is resilient to variation in, among others, conversational domain, recording equipment, recording environment, reverberation, ambient noise, number of speakers, and speaker demographics. Like its predecessors, performance is evaluated under two SAD conditions: diarization from oracle reference SAD and diarization from scratch. There are no constraints on training data, with participants allowed to use any combination of open source noise sources, number of speakers, speaker demographics, and proprietary data for system development. Recordings are sampled from $11$ diverse domains ranging from clean, near-field recordings of read audio-books to extremely noisy, highly interactive, far-field recordings of speech in restaurants to clinical interviews with children. Unlike DIHARD II, diarization from multi-channel audio is not evaluated; parties interested in this condition should instead consult the results from track 2 of CHiME-6 \cite{watanabe2020chime}, which is essentially  a rerun of the DIHARD II multichannel condition. A total of 30 organizations (forming 21 teams) from industry and academia submitted 499 valid system outputs (352 to track 1 and 147 to track 2).

In the remainder of this paper, we introduce the task, metrics, and data, as well as the baseline SAD and diarization systems. Results of the evaluation for both tracks are are reported in Section~\ref{sec:results}. 

\section{Task}
\label{sec:task}
The goal of the challenge is to automatically detect and label all speaker segments for each recording; that is: i) determine how many speakers are present; ii) for each speaker identify all corresponding speech segments. Because system  performance  is  strongly  influenced  by  the  quality  of  the speech segmentation used, two different tracks are covered:
    \begin{itemize}
        \item {\bf Track 1}  --  Diarization from reference SAD. Systems are provided with a reference speech segmentation that is generated by merging speaker turns in the reference diarization.
        \item {\bf Track 2}  --  Diarization from scratch. Systems are provided with just the raw audio input for each recording session and are responsible for producing their own speech segmentation.
    \end{itemize}

\section{Performance Metrics}
\label{sec:metrics}
As in DIHARD I and II, the primary metric is DER \cite{fiscus2006rich}, which is the sum of missed speech, false alarm speech, and speaker misclassification error rates. Because systems are provided with the reference speech segmentation for track 1, for this track DER exclusively measures speaker misclassification error. This is the metric used to rank systems on the leaderboards. For each system, we also compute a secondary metric, Jaccard error rate (JER), originally introduced for DIHARD II \cite{dihard3_eval_plan}. JER is based on the Jaccard similarity index \cite{hamers1989similarity,real1996probabilistic},
a metric commonly used to evaluate the output of image segmentation systems, which is defined as the ratio between the sizes of the intersections and unions of two sets of segments. An optimal mapping between speakers in the reference diarization and speakers in the system diarization is determined and for each pair the Jaccard index of their segmentations is computed. JER is defined as 1 minus the average of these scores, expressed as a percentage.

All metrics are computed using version 1.0.1 of the {\it dscore} tool\footnote{\dscore} without the use of forgiveness collars and with scoring of overlapped speech. For further details, please consult Section 4 of the DIHARD III evaluation plan \cite{dihard3_eval_plan} and the {\it dscore} repo.

\section{Datasets}
\label{sec:data}

\begin{table}[tbp]
    \centering
    \caption{Overview of DIHARD III datasets. The {\bf Part.} column indicates the partition (core or full), while the {\bf  \% speech} and {\bf \% overlap} columns indicate, respectively, the percentage of speech/overlapped speech in the partition.}
    \begin{tabular}{c|c|c|c|c|c}
    \hline
    Set                    & Part.    & \# rec    & \# hours       &  \% speech   &  \% overlap \\
    \hline
    \multirow{2}{*}{Dev}   & Core    & 181       & 23.94          &  78.43       & 10.04 \\
    \cline{2-6}
                           & Full    & 254       & 34.15          &  79.81       & 10.70 \\
    \hline
    \multirow{2}{*}{Eval}  & Core    & 184       & 22.73          &  77.35       &  8.75 \\
    \cline{2-6}
                           & Full    & 259       & 33.01          &  79.11       & 9.35 \\
    \hline
    \end{tabular}
    \label{tab:datasets}
\end{table}

\subsection{Overview}
The development (DEV) and evaluation (EVAL) sets consist of  selections of 5-10 minute duration samples drawn from 11 domains exhibiting wide variation in recording equipment, recording environment, ambient noise, number of speakers, and speaker demographics. These domains range in difficulty from the trivial, read audio-books recorded under clean conditions by a single speaker, to the extremely challenging, conversations between up to 6 diners recorded by a binaural microphone in restaurants with varying room acoustics and noise levels. Both adult and child speech (e.g., clinical interviews) are represented, as is speech from multiple languages (English and Chinese). For the first time, narrow-band recordings are included as well as wide-band recordings; in the narrow-band case, all recordings are drawn from the unreleased Phase II calls from the Fisher English collection conducted as part of the DARPA EARS project. All the audio recordings are distributed via LDC as $16$ kHz, mono-channel files. 

The datasets are summarized in Table~\ref{tab:datasets}. For additional details about the domains, the reader is encouraged to consult the DIHARD III evaluation plan \cite{dihard3_eval_plan}.

\subsection{Scoring partitions}
For DIHARD III, we define two partitions of the evaluation data:
    \begin{itemize}
        \item {\bf core evaluation set}  --  a ``balanced'' evaluation set in which the total duration of each domain is approximately equal
        \item {\bf full evaluation set}  --  a larger evaluation set that uses all available selections for each domain; it is a proper superset of the core evaluation set
    \end{itemize}
The core evaluation set strives for balance across domains so that the evaluation metrics are not dominated by any single domain. It mimics the evaluation set composition from DIHARD I and II. The full evaluation set includes additional material from two domains (clinical interview and CTS), potentially resulting in more stable metrics at the expense of being unbalanced. All system submissions to all tracks are scored against both sets and the results reported on the leaderboards.

\subsection{Annotation}
Reference diarization was produced by segmenting the recordings into labeled speaker turns according to the following guidelines:
\begin{itemize}
    \item split on pauses $>$ 200 ms, where a pause by speaker ``S'' is defined as any segment of time during which ``S'' is not producing a vocalization of any kind\footnote{Vocalization is defined as any noise produced by the speaker by means of the vocal apparatus; e.g., speech (including yelled and whispered speech), backchannels, filled pauses, singing, speech errors and disfluencies, laughter, coughs, breaths, lipsmacks, and humming.}
    \item attempt to place boundaries within 10 ms of the true boundary, taking care not to truncate edges of words (e.g., utterance-final fricatives or utterance initial stops)
    \item where close-talking microphones exist for each speaker, perform the segmentation separately for each speaker using their individual microphone.
\end{itemize}
Reference SAD was then derived from these segmentations by merging overlapping speech segments and removing speaker identification.

During DIHARD II, it was found that manual annotation for these specifications needed highly skilled and experienced annotators using multiple spectrogram displays, making the annotation extremely slow and expensive. Many annotators were incapable of performing the task even after extensive training and the remainder found it extremely laborious with real time rates typically greater than 15X and sometimes exceeding 30X. Consequently, for DIHARD III we abandoned a commitment to entirely manual segmentation. Where a manual segmentation to these specs already existed (i.e., files annotated for DIHARD II), we used it. For all other data we instead produced a careful turn-level transcription, then established boundaries using a Kaldi-based forced aligner.

\section{Baseline system}
\label{sec:baseline}
\subsection{Speech activity detection}
The baseline for track 2 uses a TDNN SAD model based on the Kaldi Aspire recipe (``egs/aspire/s5''). $40$-D mel frequency cepstral coefficients (MFCCs) extracted every 30 ms using a 25 ms window are fed into a neural network consisting of 5 TDNN layers \cite{peddinti2015time} followed by 2 statistics pooling layers \cite{ghahremani2016acoustic}. The network context is set to approximately 1 second (left context: 0.8 sec; right context: 0.2 sec). The DNN was trained with two classes -- speech and non-speech  --  on the DIHARD III DEV set. Training utilized the entire DEV set and was continued for 40 epochs. During inference, the posteriors of the model were converted to pseudo-likelihoods using the empirical speech/non-speech priors for the DEV set and Viterbi decoding was performed using an HMM with the following constraints: minimum speech duration: 240 ms, minimum non-speech duration: 30 ms. Miss rate, false alarm rate, and overall error (i.e., the actual frame-wise error rate) for the SAD system on the DEV and EVAL sets are depicted in Table~\ref{tab:sad_results}.

\subsection{Diarization}
The diarization baseline is based on LEAP Lab's submission to DIHARD II \cite{singh2019leap}. The system performs diarization by dividing each recording into short overlapping segments, extracting x-vectors \cite{snyder2016deep,snyder2018x}, scoring with probabilistic linear discriminant analysis (PLDA) \cite{prince2007probabilistic}, and clustering using agglomerative hierarchical clustering (AHC) \cite{han2008strategies}. The AHC ouput is then refined using variational Bayes hidden Markov model (VB-HMM) \cite{diez2018speaker,diez2019analysis} with posterior scaling \cite{singh2019leap}. The trained models and recipes for both tracks are distributed through GitHub\footnote{\dihard3baseline}.

The x-vector extractor configuration is identical to that of \cite{sell2018diarization,snyder2018x} with two exceptions: i) $30$-D MFCCs are used instead of a mel filterbank; ii) the embedding layer uses 512 dimensions. MFCCs are extracted every $10$ ms using a $25$ ms window and mean-normalized using a 3-second sliding window. For training, we use a combination of VoxCeleb 1 and VoxCeleb 2 \cite{nagrani2017voxceleb,chung2018voxceleb2} augmented with additive noise and reverberation according to the recipe from \cite{snyder2016deep}. Segments under 4-second duration are discarded, resulting in a training set with 7,323 speakers. Reverberation is added by convolution with room responses from the RIR dataset \cite{ko2017study}, while additive noises are drawn from MUSAN \cite{snyder2015musan}. At test time, x-vectors are extracted from 1.5 sec segments with a 0.25 sec shift. X-vectors are centered and whitened using statistics estimated from the DIHARD III DEV and EVAL sets, followed by length normalization \cite{garcia2011analysis} .

The x-vectors are then clustered using AHC and a similarity matrix produced by scoring with a Gaussian PLDA model \cite{prince2007probabilistic}. The PLDA model was trained using centered, whitened, and length normalized x-vectors extracted from VoxCeleb segments with duration $\geq$3 sec. Prior to PLDA scoring, dimensionality reduction was performed using conversation-dependent PCA~\cite{zhu2016online} preserving 30\% of the total variability. For each track, the stopping criteria for AHC was tuned to minimize DER on the DEV set.

%
We then refine the AHC output using frame-level VB-HMM resegmentation as described by \cite{diez2018speaker,diez2019analysis}. 24-D MFCCs are extracted every 10 ms using a 15 ms window; neither mean nor variance normalization are applied, nor do we use delta coefficients. We use a universal background model (UBM-GMM) with 1,024 diagonal-covariance components and a total variability ($\textbf{V}$) matrix containing 400 eigenvoices. Both the UBM-GMM and $\textbf{V}$ were trained using the same data as    the x-vector extractor. Following \cite{singh2019leap}, posterior scaling was applied to discourage frequent speaker transitions by the VB-HMM. This scaling was accomplished by boosting the zeroth order, but not first or second order, statistics prior to VB-HMM likelihood computation. The VB-HMM is initialized separately for each recording from the result of AHC and run for one iteration. Parameters were set to the following values by tuning on the DIHARD III DEV set: scaling factor $\beta=10$, loop probability $Ploop=0.45$, downsampling factor $downSamp=25$.

DER and JER of the baseline diarization recipe for tracks 1 and 2 are reported in Tables \ref{tab:track1_diarization_results} and \ref{tab:track2_diarization_results}, respectively.  Mirroring the findings of \cite{diez2018but,landini2020but}, VB-HMM resegmentation reliably improves DER and JER for both tracks, though the gains are more pronounced for track 2. Possibly, the effects of VB-HMM resegmentation could be enhanced by using a UBM-GMM and variablity matrix trained on or adapted to domain materials,  though we did not explore this possibility. 

\begin{table}[tbp]
    \centering
    \caption{Baseline SAD results for the core/full DEV and EVAL sets. The {\bf Part.} column indicates whether scoring was performed using the full or core DEV/EVAL set.}
    \begin{tabular}{l|l|c|c|c}
    \hline
    Set                    & Part.    &  Miss  ($\%$)   &  FA  ($\%$)   &  Overall error (\%) \\
    \hline
    \multirow{2}{*}{Dev}   & Core     & 1.84           &  3.98         &  2.30 \\
    \cline{2-5}
                           & Full     & 1.88           &  4.55         &  2.42 \\
    \hline
    \multirow{2}{*}{Eval}  & Core     & 4.97           &  15.07        &  7.26 \\
    \cline{2-5}
                           & Full     & 4.35           &  14.65        &  6.51 \\
    \hline
    \end{tabular}
    \label{tab:sad_results}
\end{table}

\begin{table}[tbp]
    \centering
    \caption{Track 1 diarization results for the core/full DEV and EVAL sets  with and without VB-HMM resegmentation.}
    \begin{tabular}{c||c|c|c|c|c}
    \hline
         Part.    & VB-HMM reseg.       &  \multicolumn{2}{c|}{DER (\%)} & \multicolumn{2}{c}{JER (\%)} \\ 
                 &                     & Dev       & Eval      & Dev      & Eval        \\ 
         \hline
         Core    & No                  & 21.05     & 21.66     & 46.34    & 48.10       \\
         \hline
         Core    & Yes                 & 20.25     & 20.65     & 46.02    & 47.74       \\
         \hline
         Full    & No                  & 20.71     & 20.75     & 42.44    & 43.31       \\
         \hline
         Full    & Yes                 & 19.41     & 19.25     & 41.66    & 42.45       \\
         \hline
    \end{tabular}
    \label{tab:track1_diarization_results}
\end{table}
\begin{table}[tbp]
    \centering
    \caption{Track 2 diarization results for the core/full DEV and EVAL sets  with and without VB-HMM resegmentation.}
    \begin{tabular}{c||c|c|c|c|c}
    \hline
         Part.    & VB-HMM reseg.       &  \multicolumn{2}{c|}{DER (\%)} & \multicolumn{2}{c}{JER (\%)} \\ 
                 &                     & Dev       & Eval      & Dev      & Eval         \\ 
         \hline
         Core    & No                  & 24.06     & 29.51     & 49.17    & 53.82        \\
         \hline
         Core    & Yes                 & 22.28     & 27.34     & 47.75    & 51.91        \\
         \hline
         Full    & No                  & 24.08     & 28.00     & 45.61    & 49.35        \\
         \hline
         Full    & Yes                 & 21.71     & 25.36     & 43.66    & 46.95       \\
         \hline
    \end{tabular}
    \label{tab:track2_diarization_results}
\end{table}

\label{sec:results}
\section{Results and Discussion}
\begin{figure*}[!tb]
    \centering
    \includegraphics[width=\linewidth]{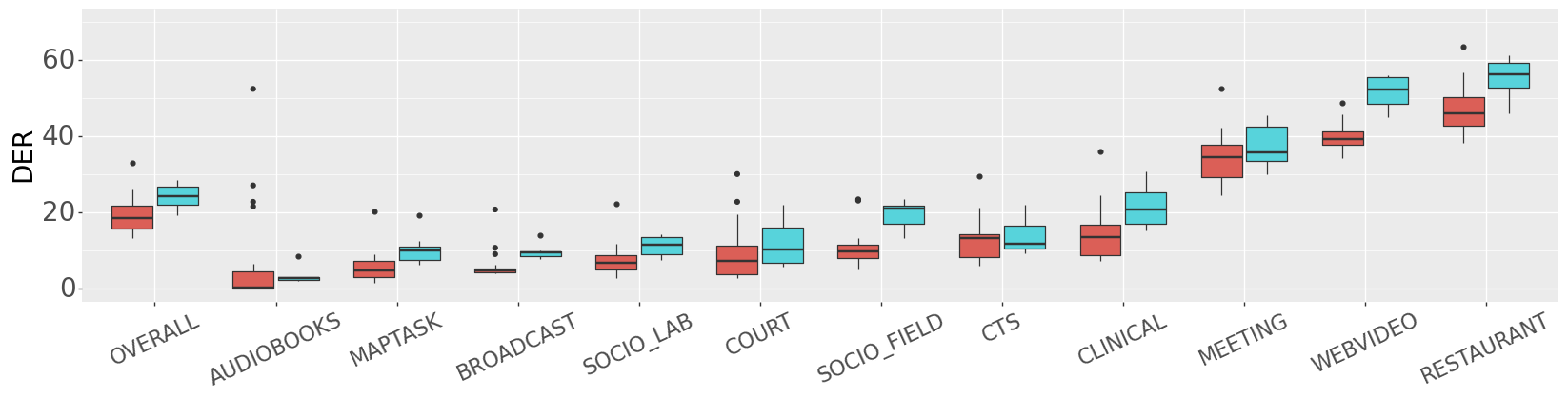}
    \caption{Track 1 (red) and track 2 (blue) DER of primary submissions by domain on the core EVAL set.}
    \label{fig:der_by_domain}
\end{figure*}

\begin{figure}[hbpt]
    \centering
    \includegraphics[width=\linewidth]{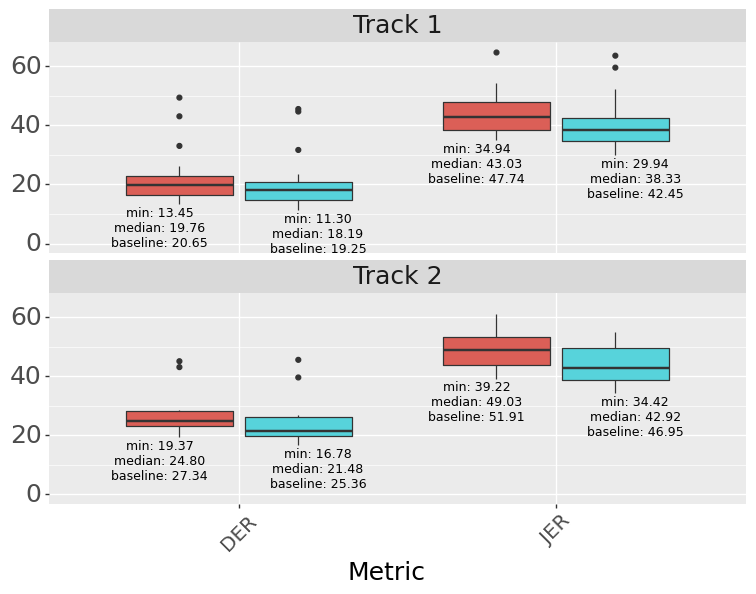}
    \caption{DER and JER of primary submissions for each track on the core (red) and full (blue) partitions of the EVAL set.}
    \label{fig:overall}
\end{figure}

Figure~\ref{fig:overall} presents boxplots of DER and JER on the core/full EVAL sets for each team's final submissions. The majority of submissions outperformed the baseline with the mean improvement in DER for these submissions being nearly 4\% absolute for track 1 and 5\% for track 2. Restricting our attention to just the top submissions, we see that the improvement is even more extreme with the winning submission \cite{wang2021ustc} outperforming the baseline by 7\% DER absolute on the core EVAL set for track 1 and 8\% on track 2. As was the case in previous DIHARD challenges, track 2 was substantially more difficult than track 1 (on the order of 5\% for both DER and JER), indicating SAD remains a challenging problem for some of the covered domains. Though the full EVAL set was more difficult than the core set, the same trends and rankings are observed for both partitions; for sake of exposition, we will report only results on the core partition throughout the rest of this discussion.

As can be seen from Figure~\ref{fig:der_by_domain}, system performance varies greatly across domains. For track 1, the median DER is below 10\% for 6 domains. With the exception of the courtroom data, these domains all consist of 1 or 2 party interactions, from which we conclude that systems are able to reliably handle clean data from a small number of speakers when a high-quality speech segmentation is available. However, in the absence of an accurate speech segmentation, for all of these domains save audiobooks DER increases substantially for all systems and catastrophically for some; particularly, for the sociolinguistic field recordings. For two of the remaining domains, clinical interviews and conversational telephone speech, DER generally ranges from 10\% to 20\% for track 1 and 15\% to 25\% for track 2, indicating that even in the two-party case there remain substantial challenges, particularly in the presence of speakers with unusual characteristics (i.e., children in the clinical interviews). For the final three domains  --  meeting speech, web videos, and restaurant  --  performance ranges awful with median track 1 DER ranging from 35\% to 45\%.

Compared to DIHARD I and II, there is a notable improvement in performance. Median DER on track 1 has fallen from over 30\% in DIHARD I, to 25\% in DIHARD II, to under 20\% in the current iteration. The improvement is even more striking for track 2, where median DER has fallen from 40\% in DIHARD I to under 25\% in DIHARD III, a 38\% relative reduction. Considering only performance of the best single system in each evaluation, DER has decreased by 43\% for track 1 (from 23.73\% to 13.45\%) and 46\% for track 2 (35.51\% to 19.37\%). These improvements are observed across all domain, including those which have not changed since DIHARD I (e.g., courtoom and meeting), indicating that these are real improvements and not an artifact of changes to the composition and annotation of the evaluation set over time. From a survey of the submitted system descriptions, these improvements appear to be due to a combination of improvements in handling of speaker overlap, use of VB-HMM in place of traditional frame-level clustering \cite{diez2018speaker,diez2019analysis}, supervised neural speech enhancement models, and target-speaker based voice activity detection (TS-VAD) \cite{medennikov2020target}, and system combination methods such as DOVER-Lap \cite{raj2021dover}. While not utilized by the winning submission, end-to-end approaches \cite{horiguchi2021hitachi} were prominent in DIHARD III and exhibited strong performance on both tracks.

\label{sec:conclusion}
\section{Summary}
We present a summary of DIHARD III, whose objective was to evaluate the current state of robust speaker diarization. Results from this present challenge indicate great progress has been made since DIHARD I, although performance of even the best system was poor for half of the domains considered, particularly when a reference SAD was not supplied. While encouraging, these results demonstrate that further work is needed to achieve the desired goal of truly robust diarization that gracefully handles the wide variety of interaction types and recording conditions observed in real world data.

\clearpage

\bibliographystyle{IEEEtran}
\bibliography{refs}

\end{document}